\documentclass[aps,prx,twocolumn,amsmath,amssymb,superscriptaddress,longbibliography]{revtex4-1}
\usepackage[english]{babel}
\usepackage{xcolor}
\usepackage{amssymb}
\usepackage{dcolumn}
\usepackage{bm}
\usepackage{graphicx}
\usepackage{amsmath}
\usepackage{braket}
\usepackage{graphicx}        % standard LaTeX graphics tool

\graphicspath{{pict/}{}}

\usepackage[normalem]{ulem}
\usepackage{dcolumn}%%%%%new
\usepackage{bm}
\usepackage[pdfstartview=FitH, CJKbookmarks=true, bookmarksnumbered=true, bookmarksopen=true, colorlinks=true, pdfborder=001, citecolor=blue, linkcolor=blue, urlcolor=blue, linktocpage=true] {hyperref}

\begin{document}
\title{Dual-mode ground-state cooling in quadratic optomechanical systems: from multistability to general dark-mode suppression}
\author{Huanhuan Wei}
\affiliation{Guangdong Provincial Key Laboratory of Quantum Metrology and Sensing $\&$ School of Physics and Astronomy, Sun Yat-Sen University (Zhuhai Campus), Zhuhai 519082, China}
\affiliation{College of Sciences/State Key Laboratory of Advanced Energy Storage Materials and Technology, Shihezi University, North Fourth Road, Shihezi City 832003, China}

\author{Yun Chen}
\affiliation{Guangdong Provincial Key Laboratory of Quantum Metrology and Sensing $\&$ School of Physics and Astronomy, Sun Yat-Sen University (Zhuhai Campus), Zhuhai 519082, China}

\author{Jing Tang}
\email{jingtang@gdut.edu.cn}
\affiliation{School of Physics and Optoelectronic Engineering, Guangdong University of Technology, Guangzhou 510006, China}
\affiliation{Guangdong Provincial Key Laboratory of Sensing Physics and System Integration Applications, Guangdong University of Technology, Guangzhou, 510006, China}

\author{Yuangang Deng}
\email{dengyg3@mail.sysu.edu.cn}
\affiliation{Guangdong Provincial Key Laboratory of Quantum Metrology and Sensing $\&$ School of Physics and Astronomy, Sun Yat-Sen University (Zhuhai Campus), Zhuhai 519082, China}

\date{\today}

\begin{abstract}
We theoretically investigate a quadratic optomechanical system comprising a single-mode optical cavity linearly coupled to one mechanical resonator and quadratically coupled to a second resonator. By tuning the cavity detuning and optomechanical coupling strengths, we demonstrate the transition from optical bistability to multistability with up to seven steady-state solutions. Notably, simultaneous ground-state cooling of both mechanical resonators occurs on the dynamically stable branch of the nonlinear steady-state solutions, offering new opportunities for combined nonlinear optical and quantum cooling functionalities. Beyond the multistable regime, we systematically study dual-mode ground-state cooling and find that robust simultaneous cooling can be achieved over a broad parameter range, except when the linear and quadratic couplings become comparable, where a dark-mode effect arises. In this case, tuning the second-order optomechanical-induced frequency shifts effectively suppresses dark-mode interference, enabling controllable and simultaneous ground-state cooling. Our results provide a versatile framework for engineering multimode quantum states in optomechanical systems and open new avenues for the development of multifunctional quantum devices, including ultra-sensitive sensors, scalable quantum memories, and integrated quantum networks.
\end{abstract}
\maketitle

\section{Introduction}

Cavity optomechanics has emerged as a transformative research frontier in quantum science, providing unprecedented opportunities to explore the quantum-classical interface and enabling a wide range of applications in quantum-enhanced metrology~\cite{science1231282}, quantum information protocols~\cite{PhysRevLett.109.013603, s41567-021-01402-0}, and ultra-sensitive quantum sensing~\cite{LiOuLeiLiu2021}.  Central to this field is the radiation-pressure-mediated interaction between optical cavity fields and mechanical motion, which has enabled the realization of macroscopic quantum phenomena~\cite{PhysRevLett.107.133601},  nonclassical state engineering~\cite{PhysRevA.88.063819}, and deciphering quantum-to-classical transitions~\cite{PhysRevA.88.023817,PhysRevLett.112.080502}.

Achieving ground-state cooling of mechanical resonators is a critical prerequisite for harnessing quantum effects in optomechanical systems~\cite{research.0206, science.1094419, PT.3.1640}. It enables the manipulation of mechanical motion at the quantum level, with applications ranging from quantum information processing~\cite{PhysRevLett.108.153603,PhysRevLett.109.130503} and high-precision control and measurement~\cite{Krause2012,LI2013223} to the study of quantum-classical boundary in macroscopic systems~\cite{PhysRevLett.109.023601,PhysRevLett.112.080502}. While extensive progress has been made in cooling single mechanical modes via cryogenic techniques~\cite{Connell2010}, feedback cooling~\cite{PhysRevLett.80.688,PhysRevLett.102.207209,Rossi2018}, and resolved sideband cooling~\cite{Teufel2011, PhysRevLett.97.243905,PhysRevLett.99.093901,PhysRevLett.99.093902,PhysRevA.77.033804,Schliesser2008,Chan2011,PhysRevLett.110.153606,PhysRevLett.116.063601,PhysRevLett.117.197202,Clark2017}, extending these approaches to simultaneously cool multiple mechanical resonators remains challenging.  A major obstacle is the presence of dark modes-mechanical modes that are decoupled from the optical cavity, which hinder efficient cooling. Developing strategies to overcome dark-mode limitations is thus a key open problem in multi-resonator optomechanics.

In parallel, optical multistability has garnered growing interest for its potential in ultra-high-speed communication, signal processing, and quantum memory architectures~\cite{10.1063,Chen2017}. As a nonlinear optical phenomenon, multistability enables systems to exhibit multiple stable states under continuous driving, offering a rich platform for exploring driven-dissipative quantum dynamics~\cite{PhysRevA.28.2569}, which provides a wide range of applications, such as all optical switches~\cite{Paraso2010,PhysRevLett.109.223906,Zhang20}, optical storage and memories~\cite{Tucker08,PhysRevLett.113.074301,Kuramochi2014,PhysRevLett.120.225301}, transistor~\cite{Ballarini2013}, filters~\cite{Zheng21}, sensors~\cite{Yang18}. Notably, extensive theoretical and experimental studies have been conducted on optical bistability (or multistability) in the quantum systems, such as optomechanical systems~\cite{PhysRevA.88.055801,PhysRevA.98.063845, Zhang2017}, cavity magnon systems~\cite{PhysRevLett.120.057202,PhysRevLett.127.183202}, atomic-cavity quantum electrodynamics (QED)~\cite{PhysRevLett.91.143904,PhysRevLett.103.160403}, semiconductor microcavities~\cite{PhysRevLett.118.247402,Fink2018} and plasmonic crystals~\cite{PhysRevLett.97.057402,Guo17}. While extensive researches on optical bistability and multistability in linear coupling systems, the interplay between quadratic coupling and multistability remains under explored. Quadratic coupling, where the optical cavity mode is coupled to the square of the mechanical displacement, introduces unique two-phonon processes that can significantly diversify system dynamics and enable novel quantum control mechanisms. Recent advances have highlighted the potential of quadratic coupling optomechanical systems to exhibit a range of fascinating phenomena~\cite{Thompson2008, Jayich_2008}, including optical multistability~\cite{Mahajan2023}, two-phonon OMIT~\cite{PhysRevA.88.013804,PhysRevA.93.043804}, photon blockade~\cite{PhysRevA.88.023853}, optical amplification~\cite{PhysRevLett.120.023601,PhysRevA.95.033803}, optical springs~\cite{PhysRevA.78.013831} and electromagnetically induced transparency (EIT)~\cite{PhysRevA.83.023823}, cooling and compression of mechanical oscillators~\cite{PhysRevLett.99.073601,PhysRevA.82.021806}. These systems offer distinct advantages, such as enhanced nonlinearity and the ability to access higher-order mechanical modes, making them a promising platform for exploring ground-state cooling and optical multistability.

\begin{figure}[t]
\includegraphics[width=0.42\textwidth]{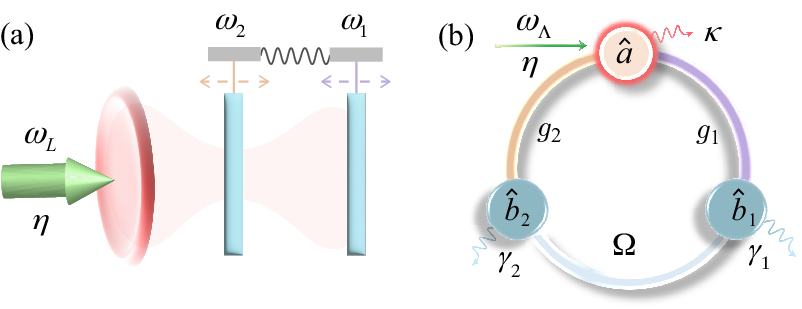}
\caption{(a) Schematic diagram of the optomechanical system. (b) A single-mode cavity mode interacts with the mechanical modes $\hat{b}_1$ and $\hat{b}_2$  via linear and quadratic couplings, respectively. Moreover, the two mechanical modes are coupled through phase-dependent phonon-tunneling interactions, characterized by a coupling strength  $\Omega$ and phase $\theta$.}
\label{fig:model}
\end{figure}

Motivated by these opportunities, we theoretically investigate a quadratic optomechanical system comprising a single-mode optical cavity linearly coupled to one mechanical resonator and quadratically coupled to a second resonator. We first demonstrate that tuning the cavity detuning and coupling strengths induces transitions from bistability to multistability with up to seven steady-state solutions. Intriguingly, we observe that ground-state cooling of both mechanical resonators can occur on the dynamically stable branch within the nonlinear regime, suggesting a novel interplay between nonlinear steady-state structure and quantum cooling. Building on this observation, we then systematically explore dual-mode ground-state cooling beyond the multistable regime. In particular, we identify second-order optomechanical-induced frequency shifts as an effective mechanism to suppress dark-mode interference and enable robust, controllable simultaneous cooling across a broad parameter space. Our findings provide a comprehensive framework for engineering multimode quantum states in optomechanical systems and open new avenues for multifunctional quantum devices based on quadratic coupling, such as ultra-sensitive sensors\cite{2023Entanglement}, quantum memories\cite{PhysRevA.96.013854}, and integrated quantum computing architectures\cite{Clerk2020}.

\section{Model and Hamiltonian}
We consider an optomechanical system comprising an optical cavity field and two coupled mechanical modes. The cavity field interacts linearly with mechanical oscillator $\hat{b}_1$ and quadratically with mechanical oscillator $\hat{b}_2$, and the two mechanical modes are coupled via a phase-dependent phonon-exchange interaction with coupling strength $\Omega$, as illustrated in Fig.~\ref{fig:model}. Additionally, an external driving field with frequency $\omega_{L}$ and amplitude $\eta$ is applied to the optical cavity.
Under the rotating-wave approximation, the relevant Hamiltonian for the system reads:
\begin{eqnarray}
\hat{H}/\hbar &=  \Delta_{c}\hat{a}^{\dagger}\hat{a}+\omega_{1}\hat{b}_{1}^{\dagger}\hat{b}_{1}+\omega_{2}\hat{b}_{2}^{\dagger}\hat{b}_{2}+g_{1}\hat{a}^{\dagger}\hat{a}(\hat{b}_{1}^{\dagger}+\hat{b}_{1})
\nonumber \\
 &  +g_{2}\hat{a}^{\dagger}\hat{a}(\hat{b}_{2}^{\dagger}+\hat{b}_{2})^{2}+[(\Omega e^{i\theta}\hat{b}_{1}^{\dagger}\hat{b}_{2}+\eta \hat{a})+\rm H.c.],
 \label{eq:H2}
\end{eqnarray}
where $\Delta_{c} = \omega_{c} - \omega_{L}$ represents the cavity-light detuning of the system, with $\omega_{c}$ being the bare cavity frequency.  The operators $\hat{a}^{\dagger}$ ($\hat{a}$) and $\hat{b}^{\dagger}_l$ ($\hat{b}_l$) (for $l=1,2$) denote the creation (annihilation) operators of the cavity field and the mechanical modes, respectively. $\omega_{1}$ and $\omega_{2}$ are frequencies of the mechanical modes. $g_{1}$ and $g_{2}$ correspond to the linear and quadratic optomechanical coupling strengths for the cavity mode and the mechanical resonators.  To facilitate energy exchange between the two mechanical resonators, a phase-dependent phonon-exchange interaction with coupling strength $\Omega$ and phase $\theta$ is introduced. We note that this phase-dependent phonon-exchange interaction $\Omega e^{i\theta}$ represents an independently engineered coherent coupling between the two mechanical modes (e.g., mediated by an auxiliary optical field or an external modulation scheme), whose amplitude and phase are controlled by external drives. This interaction is introduced independently of the quadratic optomechanical interaction, which instead modifies the steady-state configuration and effective parameters in the linearized dynamics.

By incorporating dissipation and noise terms into the Heisenberg equations of motion, $i\hbar \frac{d\hat{O} }{dt}=[\hat{O},\hat{H}]$, based on the Hamiltonian~(\ref{eq:H2}),
the corresponding Langevin equations for the annihilation operators of the cavity and mechanical modes are given by:
\begin{eqnarray}
\dot{\hat{a}} & = & -\left\{ \kappa+i\left[\Delta_{c}+g_{1}(\hat{b}_{1}^{\dagger}+\hat{b}_{1})+g_{2}(\hat{b}_{2}^{\dagger}+\hat{b}_{2})^{2}\right]\right\}\hat{a}
\nonumber \\
 &  &-i\eta^{*}+\sqrt{2\kappa}\hat{a}_{in},\nonumber \\
\dot{\hat{b}}_{1} & = & -(\gamma_{1}+i\omega_{1})\hat{b}_{1}-ig_{1}\hat{a}^{\dagger}\hat{a}-i\Omega e^{i\theta}\hat{b}_{2}
+\sqrt{2\gamma_{1}}\hat{b}_{1,in},\nonumber \\
\dot{\hat{b}}_{2} & = & -(\gamma_{2}+i\omega_{2})\hat{b}_{2}-2i({\hat{b}_{2}}^{\dagger}+\hat{b}_{2})g_{2}\hat{a}^{\dagger}\hat{a}
-i\Omega e^{-i\theta}\hat{b}_{1}\nonumber \\&  &+\sqrt{2\gamma_{2}}\hat{b}_{2,in}.
\label{eq:langevin}
\end{eqnarray}
Here, $\kappa$ is the decay rate of the optical cavity, and $\gamma_{l}$ are the damping rates of the mechanical modes. The operators $\hat{a}_{in}$ ($\hat{a}_{in}^{\dagger}$) and $\hat{b}_{l,in}$ ($\hat{b}_{l,in}^{\dagger}$) represent the noise operators associated with the cavity and mechanical modes, respectively. These noise operators have zero mean values, and their correlations satisfy the following relations:
\begin{eqnarray}
\left\langle \hat{a}_{in}(t)\hat{a}_{in}^{\dagger}(t')\right\rangle  & = & \delta(t-t'),\nonumber\\
\left\langle \hat{a}_{in}^{\dagger}(t)\hat{a}_{in}(t')\right\rangle  & = & 0,\nonumber\\
\left\langle \hat{b}_{l,in}(t)\hat{b}_{l,in}^{\dagger}(t')\right\rangle  & = & (\bar{n}_{l}+1)\delta(t-t'),\nonumber\\
\left\langle \hat{b}_{l,in}^{\dagger}(t)\hat{b}_{l,in}(t')\right\rangle  & = & \bar{n}_{l}\delta(t-t'),
\end{eqnarray}
where $\bar{n}_{l}$ represents the mean thermal phonon occupancy associated with the thermal bath of the mechanical resonators. The optical cavity mode interacts with a vacuum bath, while the mechanical resonators are coupled to thermal baths characterized by $\bar{n}_{l}$.
The cooling mechanism of the mechanical resonators arises from the interaction with the vacuum bath of the cavity field, which effectively extracts thermal excitations from the mechanical modes, thereby serving as a cooling reservoir.
\begin{figure}[t]
\includegraphics[width=0.48\textwidth]{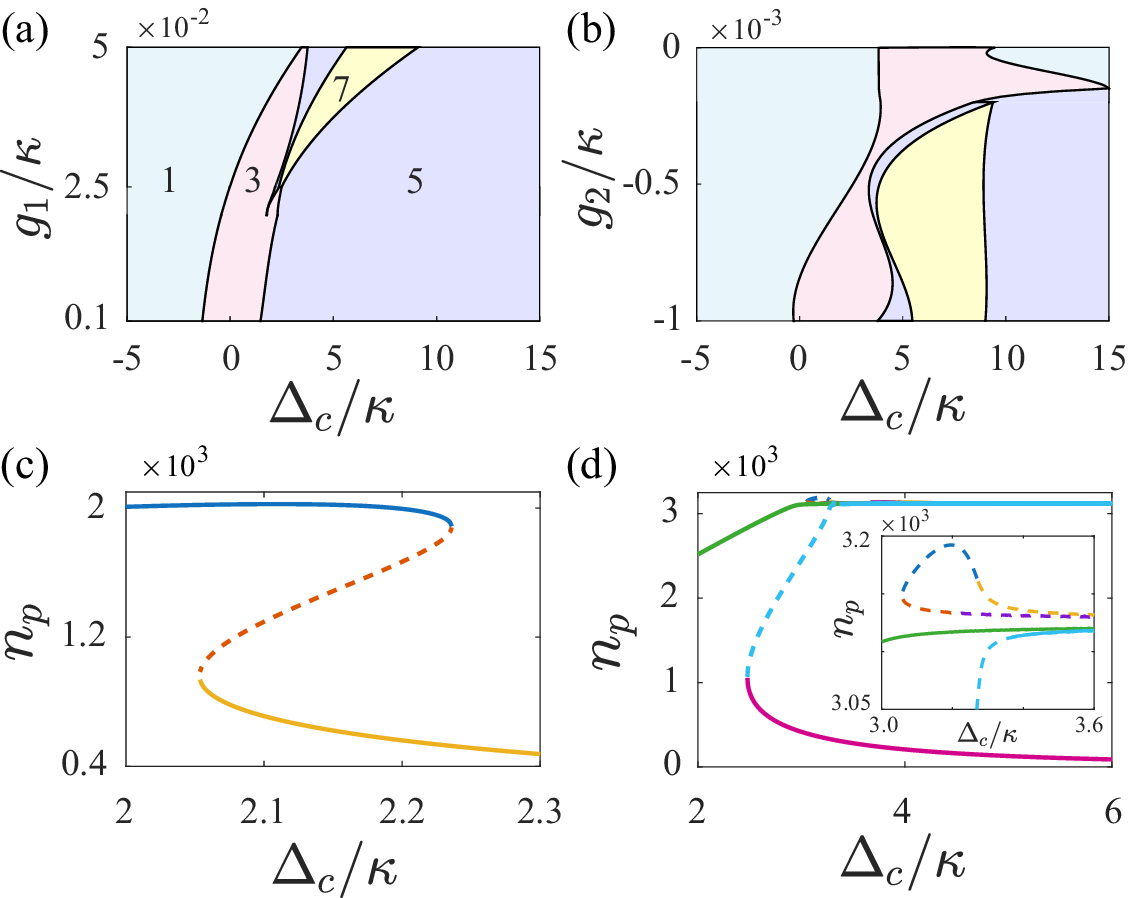}
\caption{(a) Phase diagram in the $g_1-\Delta_c$ plane with $g_2/\kappa=-0.0004$, $\eta/\kappa=95$ and $\Omega/\kappa=1$. The numbers (1,3,5,7) labeled in the phase diagrams indicate the number of real solutions to the steady-state algebraic equation, while their dynamical stability is analyzed separately. (b) Phase diagram in the $g_2-\Delta_c$ plane with $g_1/\kappa=0.05$, $\eta/\kappa=95$ and $\Omega/\kappa=1$. (c) Intracavity photon number $n_p$ as a function of $\Delta_c$ for steady-state solutions with $g_1/\kappa=0.05$, $g_2/\kappa=0$, $\eta/\kappa=45$ and $\Omega/\kappa=1$. (d) $n_p$ versus $\Delta_c$  with $g_1/\kappa=0.05$, $g_2/\kappa=-0.0004$, $\eta/\kappa=56.5$ and $\Omega/\kappa=0.005$. The other parameters are $\omega_1/\kappa=\omega_2/\kappa=5$ and $\theta=\pi$. }
\label{fig:3w}
\end{figure}
As to the experimental feasibility, the linear optomechanical coupling strength $g_1$ has been widely implemented in well-established  platforms such as micro-ring/disk resonators~\cite{Schliesser2008}, optomechanical crystal nanobeams~\cite{Chan2011}, and membrane-in-the-middle (MIM) systems~\cite{Thompson2008, Jayich_2008}. The crucial quadratic optomechanical coupling strength $g_2$ can be achieved by precisely positioning a membrane at the node or antinode of the optical standing wave field~\cite{Thompson2008, Jayich_2008, PhysRevA.82.021806}. This approach has already been employed in MIM systems to observe phenomena such as optical bistability~\cite{Mahajan2023} and two-phonon cooling~\cite{PhysRevA.82.021806}. Furthermore, the phase-dependent phonon-exchange interaction $\Omega e^{i\theta}$, which is essential for achieving cooperative quantum control, can be implemented by using an auxiliary optical field that  mediates  coupling between the two mechanical modes~\cite{PhysRevA.102.011502}. By adjusting the relative phase $\theta$ between this auxiliary field and the main cooling laser, a synthetic gauge potential can be introduced, a technique that has already been experimentally demonstrated in nonreciprocal ground-state cooling systems.

In our numerical simulations, experimentally accessible parameters were used for both the multistability and ground-state cooling analyses. For multistability, the cavity decay rate was set to $\kappa = 2\pi\times500$ MHz as the characteristic energy scale~\cite{CH2011}, with mechanical frequencies $\omega_1/\kappa = \omega_2/\kappa = 5$. Owing to the high mechanical quality factors achievable in experiments, damping was neglected in this section. For ground-state cooling, the mechanical frequency $\omega_1 = 2\pi\times3.68$ GHz was used as the energy unit~\cite{CH2011}, with damping rates $\gamma_{1,2}/\omega_{1,2} = 2\times10^{-6}$ and initial mean thermal phonon numbers $\bar{n}_{1,2} = 300$, corresponding to a cryogenic environment of  about 10 mK attainable in dilution refrigerators. The tunable parameters across both analyses include the linear and quadratic coupling strengths, optical detuning ($\Delta_c$, $\Delta$), driving strength $\eta$ phonon-exchange interaction $\Omega$, phase $\theta$, and quadratic optomechanical frequency shift $G_{22}$. 

\section{Optical multistability}
For the quadratic optomechanical system described above, we decompose the three modes ($\hat{a}$ and $\hat{b}_{l=1,2}$) into their steady-state components and fluctuations, i.e. ($\hat{a}=\alpha+\delta \hat{a}$, $\hat{b}_{1}=\beta_{1}+\delta \hat{b}_{1}$ and $\hat{b}_{2}=\beta_{2}+\delta \hat{b}_{2}$). Then the  system's dynamics, governed by the Hamiltonian~(\ref{eq:H2}), can then be expressed through the Heisenberg equations of motion:
\begin{eqnarray}
\dot{\alpha} & = & -(\kappa+i\Delta)\alpha-i\eta^{*}\nonumber \\
\dot{\beta_{1}} & = & -(\gamma_{1}+i\omega_{1})\beta_{1}-ig_{1}\left|\alpha\right|^{2}-i\Omega e^{i\theta}\beta_{2}\nonumber \\
\dot{\beta_{2}} & = & -(\gamma_{2}+i\omega_{2})\beta_{2}-2ig_{2}\beta_{2}\left|\alpha\right|^{2}-2ig_{2}\beta_{2}^{*}\left|\alpha\right|^{2}\nonumber \\&&-i\Omega e^{-i\theta}\beta_{1}.
\label{eq:dot13}
\end{eqnarray}
Here $\Delta=\Delta_{c}+2g_{1}{\rm }Re[\beta_{1}]+g_{2}(\beta_{2}^{*2}+\beta_{2}^{2}+2\left|\beta_{2}\right|^{2})$ represents the normalized driving detuning of the cavity field. Then the steady-state solutions of the classical equations of motion (where the time derivatives vanish, ignoring decay terms) are
\begin{eqnarray}
\alpha & = & \frac{-i\eta^{*}}{\kappa+i \Delta}\nonumber\\
\beta_{1} & = & \frac{i(g_{1}\left|\alpha\right|^{2}+\Omega e^{i\theta}\beta_{2})}{-i\omega_{1}}\nonumber\\
\beta_{2} & = & \frac{i(2g_{2}{\rm }\beta_{2}^{*}\left|\alpha\right|^{2}+\Omega e^{-i\theta}\beta_{1})}{-i\omega_{2}-2ig_{2}\left|\alpha\right|^{2}}.
\end{eqnarray}
Building upon the above equations, we derive the  condition for optical multistability through a systematic algebraic analysis. The steady-state intracavity photon number $n_{p}=\left|\alpha\right|^{2}$ satisfies the following seventh-order polynomial equation:
\begin{align}
\sum \limits _{m=0} ^{7}C_{m}n_{p}^{m} =0.
\label{eq:c7}
\end{align}
where the coefficients $C_m$ are explicitly defined as
\begin{widetext}
\begin{eqnarray} \label{c1c7}
C_{7} & = &64g_{1}^4 g_{2}^4\omega_{1}^2\{4(x+\omega_{1} \omega_{2})[(x+\omega_{1} \omega_{2})-\Omega ^2 y]+\Omega ^4 y^2 \},\nonumber \\
C_{6} & = &32g_{1}^4g_{2}^3\omega_{1}x \{[2(x+2\omega_{1}\omega_{2})]^2-16\omega_{1}^2\omega_{2}^2
-\Omega ^2[ 2(1+\omega_{1}\omega_{2})+y(4+9\omega_{1}\omega_{2})]+\frac{3}{2} \Omega ^4 y^2 \}\nonumber\\&  &-256g_{2}^4g_{1}^2\omega_{1}^3 x\Delta_{c}\{2(x+\omega_{1} \omega_{2})+\Omega ^2 y\},\nonumber \\
C_{5} & = &16g_{1}^4 g_{2}^2x^2\{(x+\sqrt{5}\omega_{1}\omega_{2})^2+8\omega_{1}^2\omega_{2}^2
+[\frac{3}{4}\Omega ^4-\frac{1}{2}\Omega ^2(3x+13\omega_{1}\omega_{2})]y+\Omega ^4[\frac{11}{8}+\frac{9}{16}(y^2-2)] -\Omega ^2 (x+3\omega_{1}\omega_{2})\} \nonumber \\
 &  &+256g_{2}^4\omega_{1}^4 x^2z+64g_{2}^3
g_{1}^2 \omega_{1}^2 x^2\Delta_{c}[ \Omega ^2+\frac{7}{2}\Omega ^2y-(6x+10\omega_{1}\omega_{2})],\nonumber \\
C_{4} & = & 16 g_{1}^4 g_{2}x^3\omega_{2}[x+3 \omega_{1}\omega_{2}-\frac{1}{2} \Omega ^2-\frac{3}{4} \Omega ^2y]
+32g_{1}^2g_{2}^2 x^3\omega_{1} \Delta_{c}[-9\omega_{1}\omega_{2}-3x+\Omega ^2+2\Omega ^2y]
-256 [g_{2}^4 \omega_{1}^4 x^2  \eta^2- g_{2}^3 x^3\omega_{1}^3 z], \nonumber\\
C_{3} & = &-4 x^4 g_{1}^2 g_{2}\Delta_{c}[2x+14 \omega_{1} \omega_{2}- \Omega ^2-\frac{3}{2} \Omega ^2y]
+4 g_{1}^4 x^4 \omega_{2}^2-256 g_{2}^3 x^3 \omega_{1}^3 \eta^2
+96 g_{2}^2 x^4 \omega_{1}^2z,\nonumber \\
C_{2} & = &-4 g_{1}^2 x^5 \omega_{2}\Delta_{c}-96 g_{2}^2 x^4 \omega_{1}^2 \eta^2+16 g_{2}x^5 \omega_{1}z,\nonumber \\
C_{1} & = &-16 g_2 x^5 \eta^2 \omega_{1}+x^6  {z},\nonumber \\
C_{0} & = &-\eta^{2}x^{6}.
\end{eqnarray}
\end{widetext}
here $x=\omega_{1}\omega_{2}-\Omega^{2}, y=e^{ i 2\theta}+e^{-i2\theta}$, and $z=\Delta_{c}^2+ \kappa^2$. 

Equation~(\ref{eq:c7}) provides a compact representation of the nonlinear steady-state condition, where the coefficients  $C_m$ explicitly depend on the linear and quadratic optomechanical coupling strengths $g_{1}$ and $g_{2}$. These coefficients determine the number and stability of the steady-state solutions, enabling a comprehensive understanding of the system's multistable behavior.

In the absence of optomechanical couplings ($g_{1}$=$g_{2}= 0$), Eq.~(\ref{eq:c7}) reduces to a linear response with a single steady-state solution. Once optomechanical interactions are introduced ($g_1 \neq 0$ and/or $g_2 \neq 0$), the system's behavior changes qualitatively, and multiple steady states, both stable and unstable can emerge. 
This clearly highlights the central role of optomechanical nonlinearities in enabling optical bistability and multistability. In particular,  when only linear coupling is present ($g_1 \neq 0$, $g_2 = 0$), Eq.~(\ref{eq:c7}) reduces to a third-order polynomial, then the system exhibits three steady-state solutions, corresponding to a conventional optical bistability scenario. Physically, this implies that the inclusion of linear coupling modifies the optical response, introducing nonlinear shifts that lead to a hysteresis-like behavior in the cavity field dynamics. When only quadratic coupling is present($g_1 = 0$, $g_2 \neq 0$), the system exhibits five distinct steady-state solutions. This stems from the fact that Eq.~(\ref{eq:c7}) now takes the form of a fifth-order polynomial, which allows for additional solutions compared to the linear coupling case. The emergence of five roots suggests that quadratic coupling alone can induce higher-order multistability. This is likely due to its ability to introduce asymmetric potential modifications and enhance nonlinear feedback effects, which significantly alter the cavity dynamics. Furthermore, when both linear and quadratic couplings are simultaneously present, their interaction further enriches the phase space, leading to the emergence of higher-order steady-state solutions, as illustrated in Fig.~\ref{fig:3w}. The combined effect of $g_1$ and $g_2$ suggests a highly tunable mechanism for controlling multistability, where the relative strengths of the two couplings determine the number and stability of solutions. This tunability has significant implications for applications in optical switching, quantum memory, and nonlinear photonics, where precise control over multistability is essential.

To investigate the influence of both linear and quadratic optomechanical couplings on the multistability characteristics of the system, we first examine the steady-state solutions in the $g_1$-$\Delta_c$ and $g_2$-$\Delta_c$ parameter planes, as shown in Figs.~\ref{fig:3w}(a) and \ref{fig:3w}(b). The regions labeled 1, 3, 5, and 7 denote the total number of real solutions of the nonlinear steady-state equation governing the intracavity photon number. We note that this count corresponds to the total number of real algebraic steady-state solutions and does not necessarily imply dynamical stability. These multiple solutions originate from the nonlinear self-consistent equation, whose polynomial order is determined by the combined effects of linear and quadratic optomechanical couplings. In Fig.~\ref{fig:3w}(a), the role of linear coupling strength $g_1$ is analyzed as a function of cavity detuning $\Delta_c$, with the quadratic coupling fixed at $g_2/\kappa=-0.0004$. When the linear coupling is absent ($g_1=0$), the steady-state equation reduces to a higher-order nonlinear polynomial solely determined by the quadratic interaction, yielding up to five real steady-state solutions as $\Delta_c$ varies, as predicted in previous theoretical studies.  As $g_1$ increases, the interplay of linear and quadratic radiation-pressure forces modifies the nonlinear equation. For small values of $g_1$, a single steady-state dominates over a broad detuning range. With increasing $\Delta_c$, additional real roots emerge. For sufficiently large $g_1$, up to seven real steady-state solutions of the algebraic equation exist, reflecting the enhanced effective nonlinearity arising from the combined couplings. This systematic increase in solution multiplicity indicates the strengthening of the effective nonlinear response induced by the interplay of the two coupling mechanisms.

In contrast, Fig.~\ref{fig:3w}(b) illustrates the dependence on quadratic coupling  $g_2$ with finite linear interaction ($g_1/\kappa=0.05$). For small $|g_2|$, the system exhibits at most three steady-state solutions, characteristic of conventional bistability.  As $|g_2|$ increases, the nonlinear contribution from the quadratic interaction becomes increasingly significant, effectively increasing the order of the steady-state polynomial equation and reshaping the effective potential landscape, leading to five and eventually seven steady-state solutions. This demonstrates that the quadratic coupling plays a crucial role in enhancing nonlinear interactions and enabling higher-order multistability beyond conventional bistability.

We emphasize that multiple algebraic solutions do not necessarily imply genuine multistability. A steady-state is physically relevant only if it is dynamically stable, determined by the eigenvalues of the corresponding Jacobian matrix: solutions are stable when all eigenvalues have negative real parts and unstable otherwise.  In certain regions of Figs.~\ref{fig:3w}(a) and \ref{fig:3w}(b), although multiple solutions coexist, stability analysis reveals that only the lowest branch remains stable, while the others are unstable. In this case, the system exhibits mathematical multivaluedness rather than true dynamical multistability.

To explicitly identify the parameter regimes supporting genuine multistability, defined as the coexistence of multiple dynamically stable steady states, we further examine the detuning dependence of the intracavity photon number $n_p$, as shown in Figs.~\ref{fig:3w}(c) and \ref{fig:3w}(d). Stable (unstable) solutions are indicated by solid (dashed) curves. Fig.~\ref{fig:3w}(c) corresponds to the case without quadratic coupling ($g_2=0$), where the linear coupling induces a cubic nonlinearity, leading to conventional bistability: two stable branches coexist within a finite detuning window, separated by an unstable branch. When quadratic coupling is introduced [Fig.~\ref{fig:3w}(d)], the nonlinear structure is modified, showing more complex characteristics, with single, three and five steady-states emerging as $\Delta_c$ varies. The interplay between $g_1$ and $g_2$  increases the effective nonlinear order and reshapes the potential landscape, allowing additional stable branches to emerge. In this regime, multiple dynamically stable steady states coexist over a finite detuning range, demonstrating genuine higher-order multistability. These results clarify that while linear coupling supports conventional bistability, quadratic coupling is essential for stabilizing higher-order multistability. The coexistence and competition between $g_1$ and $g_2$ therefore provide a controllable mechanism for engineering complex multistable optical states.

\begin{figure}[t]
\includegraphics[width=0.48\textwidth]{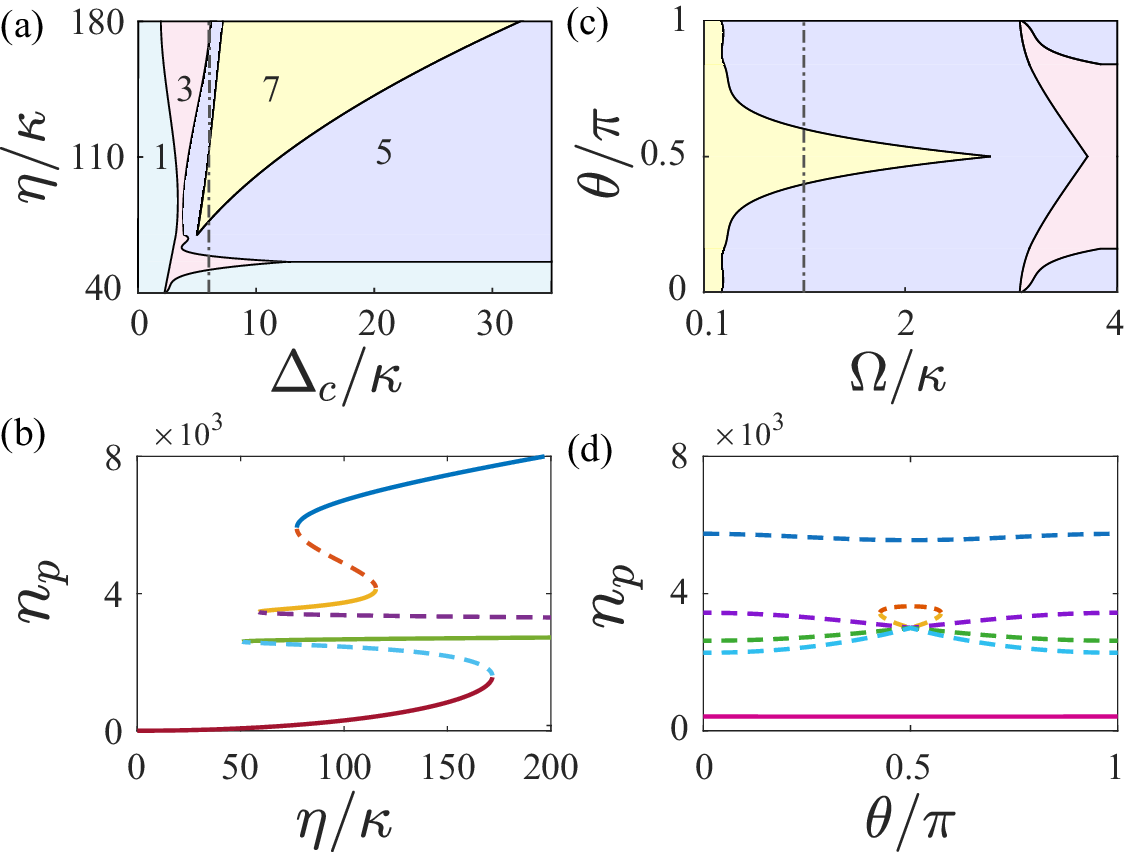}
\caption{(a) Phase diagram of the system on the $\eta-\Delta_c$ parameter plane with $\Omega/\kappa=1$ and $\theta=\pi$. (b) Intracavity photon numbers $n_p$ versus $\eta/\kappa$ with $\Delta_c/\kappa=6$, corresponding to the  dashed line in (a).
(c) Phase diagram of the system on the $\Omega-\theta$ parameter plane with $\eta/\kappa=95$ and $\Delta_c/\kappa=5$.
(d) $n_p$ versus $\theta/\pi$ for steady-state solutions with $\Delta_c/\kappa=5$ and $\Omega/\kappa=1$, corresponding to the  dashed line in (c). The other parameters are $\omega_1/\kappa=\omega_2/\kappa=5$, $g_{1}/\kappa=0.05$ and $g_{2}/\kappa=-0.0004$.}
\label{fig:4_eta1}
\end{figure}
Beyond the influence of optomechanical coupling strengths, the driving light intensity $\eta$ introduces an additional degree of control over multistability transitions. Previous discussions focused on how $g_1$ and $g_2$ shape the system's nonlinear response, here we explore how external driving strength interacts with these couplings to further enhance nonlinearity and modify stability landscapes. To illustrate this effect, Fig.~\ref{fig:4_eta1}(a) maps the evolution of multistability in the ($\eta$, $\Delta_c$) parameter space, with fixed phonon-exchange parameters $\Omega/\kappa=1$ and $\theta=\pi$ and optomechanical couplings set at $g_{1}/\kappa=0.05$ and  $g_{2}/\kappa=-0.0004$.  At low driving intensities, the system remains in a monostable state, exhibiting a predominantly linear optical response. As $\eta$ increases, nonlinear effects become significant, leading to bistability and bistability and tristability (with five steady-state solutions) as $\Delta_c$ varies. This suggests that the increased photon population enhances optomechanical interactions, allowing the system to access multiple stability branches. Further increasing $\eta$ enables higher-order multistability with up to seven coexisting steady-state solutions. This result demonstrates that optomechanical nonlinearities become dominant at high driving strengths, allowing the system to access multiple stability branches. This trend highlights the increasing role of optomechanical nonlinearities, where higher-order coupling effects become dominant, underscoring the ability of driving strength to tune and expand the accessible multistability domains, which could be useful for optical switching applications.  

To further investigate the role of driving strength, Fig.~\ref{fig:4_eta1}(b) plots the intracavity photon number $n_p$ as a function of $\eta$ at a fixed detuning $\Delta_c/\kappa = 6$. The strong dependence of $n_p$ on $\eta$ confirms that increasing the driving strength enhances the underlying nonlinear interactions, giving rise to a rich and complex multistable structure in the presence of quadratic optomechanical coupling. In addition, among the seven steady-state branches, four correspond to dynamically stable solutions (solid lines), while the remaining three are unstable (dashed lines). The system initially follows the lowest stable branch (red). As the pump strength increases, it undergoes successive transitions to the second (green), third (orange), and eventually the uppermost stable branch (blue). It is noted that the full set of algebraic steady-state solutions forms continuous branches when extended over a broader range of $\eta$, while the apparent discontinuities in the present plot arise from the limited parameter range chosen to highlight the multistable structure. The observed transitions correspond to switching between dynamically stable branches. These transitions demonstrate that, when both linear and quadratic optomechanical couplings are present, the system can sustain four distinct stable steady-state photon-number branches, highlighting the high tunability of the nonlinear optical response.

Phonon interactions $\Omega$ and the phase $\theta$ also play crucial roles in governing multistability transitions, as illustrated in Fig.~\ref{fig:4_eta1}(c) with $\eta/\kappa=95$. The phase diagram in the ($\Omega$, $\theta$) parameter space reveals a mirror symmetry around $\theta$=$\pi/2$, indicating that multistability evolution follows equivalent patterns across symmetric phase domains. This symmetry is evident both in the phase-space distributions and in the dependence of multistability on $\Omega$. Under weak phonon coupling (small $\Omega$), the system sustains up to seven coexisting steady-state solutions, suggesting that in weak phonon interaction regimes, the optomechanical interactions remain complex and supports multiple stability branches. As $\Omega$ increases,  the system undergoes successive stability reductions in the number of stable steady-state solutions,  transitioning from a highly multistable regime with seven coexisting solutions to a five-state configuration, and eventually to a bistable state. This trend suggests that stronger phonon interactions effectively suppress higher-order multistability, simplifying the system's dynamical behavior and leading to fewer stable solutions with a more predictable response. At even stronger phonon interactions, the system converges to a three-stable-solutions configuration. This indicates that increasing coupling strength suppresses higher-order multistability and reduces the number of coexisting steady-state branches, thereby simplifying the steady-state structure.

Fig.~\ref{fig:4_eta1}(d) shows the multistability behavior of photon numbers as a function of phase $\theta$ for $\Delta_c/\kappa=5$ and $\Omega/\kappa=1$. The system exhibits a clear mirror symmetry about $\theta=\pi/2$, with the multistability evolution on both sides following nearly identical trends. This mirror symmetry originates from the mathematical structure of the steady-state equation Eq.~(\ref{eq:c7}), whose coefficients $C_m$ in Eq.~(\ref{c1c7}) depend on the phase $\theta$ only through the composite term $y = e^{i2\theta} + e^{-i2\theta} = 2\cos(2\theta)$, which is invariant under the transformation $\theta \to \pi-\theta$. Physically, this invariance reflects that the phonon-exchange interaction $\Omega e^{i\theta}$ affects the system dynamics only through the relative phase between the two mechanical modes. Reversing the phase about $\pi/2$ leaves this relative interference unchanged, leading to identical steady-state responses on both sides of the symmetry axis. Specifically, within the interval $0<\theta<\pi/2$ ($\pi/2<\theta<\pi$), the system supports transitions from multistability with five-solutions to seven-solutions, consistent with the previously observed behavior in the ($\Omega, \theta$) parameter space. In this regime, stability analysis shows that the upper branches lose dynamical stability as the corresponding eigenvalues of the Jacobian matrix acquire positive real parts, while only the lowest branch remains dynamically stable. This indicates that the existence of multiple algebraic solutions does not necessarily imply multiple stable steady states. These findings reveal that, in the presence of optomechanical couplings and phase-dependent phonon interactions, the driving strength $\eta$, phonon interactions $\Omega$ and phase $\theta$ act as control parameters, enabling the tuning of multistability behavior and the emergence of complex steady-state landscapes. This has potential applications in optical switching, nonlinear photonics, and optomechanical memory devices, where controlled multistability can be leveraged for state storage and retrieval.

\section{Dual-Mode Ground-State Cooling on Dynamically Stable Branches}
Having established the nonlinear steady-state structure and the existence of genuine optical multistability in certain parameter regimes, we now turn to the problem of dual-mode ground-state cooling. Ground-state cooling is a fundamental prerequisite for realizing quantum control of macroscopic mechanical systems, enabling applications in quantum information processing, precision metrology, and fundamental tests of quantum mechanics. The interplay between optical and mechanical degrees of freedom not only governs higher-order nonlinear steady-state structure but also provides a pathway for extracting thermal excitations from mechanical resonators.  It is essential to emphasize that the cooling analysis relies on linearization around a semiclassical steady state. Therefore, only dynamically stable steady-state solutions are physically meaningful for cooling. Throughout this section, we strictly restrict our analysis to such stable branches.

To achieve efficient cooling, we consider a strongly driven optical field, ensuring that the intracavity photon number is sufficiently large to justify a linearized treatment of the system's dynamics. Under this condition, the nonlinear optomechanical interaction can be approximated by an effective linear coupling, simplifying the analysis while preserving the essential physics of quantum fluctuations. The linearized equations of motion for quantum fluctuations, derived from the quantum Langevin Eq~(\ref{eq:langevin}), can be expressed as follows:

\begin{eqnarray}
\delta\dot{\hat{a}} & = & -(\kappa+i\Delta)\delta \hat{a}-iG_{1}(\delta \hat{b}_{1}^{\dagger}+\delta \hat{b}_{1})-iG_{2}(\delta \hat{b}_{2}^{\dagger}+\delta \hat{b}_{2})
\nonumber \\
 &  &+\sqrt{2\kappa} \delta \hat{a}_{in},\nonumber\\
\delta\dot{\hat{b}}_{1} & = & -(\gamma_{1}+i\omega_{1})\delta \hat{b}_{1}-iG_{1}^{*}\delta \hat{a}-iG_{1}\delta \hat{a}^{\dagger}-i\Omega e^{i\theta}\delta \hat{b}_{2}
\nonumber \\
 &  &+\sqrt{2\gamma_{1}}\delta \hat{b}_{1,in},\nonumber\\
\delta\dot{\hat{b}}_{2} & = & -(\gamma_{2}+i\tilde{\omega}_{2})\delta \hat{b}_{2}-2iG_{22}\delta \hat{b}_{2}^{\dagger}-iG_{2}\delta \hat{a}^{\dagger}-iG_{2}^{*}\delta \hat{a}
\nonumber \\
 &  &-i\Omega e^{-i\theta}\delta \hat{b}_{1}
 +\sqrt{2\gamma_{2}}\delta \hat{b}_{2,in},\nonumber\\
\delta\dot{\hat{a}}^{\dagger} & = & -(\kappa-i\Delta)\delta \hat{a}^{\dagger}+iG_{1}^{*}(\delta \hat{b}_{1}^{\dagger}+\delta \hat{b}_{1})+iG_{2}^{*}(\delta \hat{b}_{2}^{\dagger}+\delta \hat{b}_{2})
\nonumber \\
 &  &+\sqrt{2\kappa} \delta \hat{a}_{in}^{\dagger},\nonumber\\
\delta\dot{\hat{b}}_{1} ^{\dagger}& = & -(\gamma_{1}-i\omega_{1})\delta \hat{b}_{1}^{\dagger}+iG_{1}\delta \hat{a}^{\dagger}+iG_{1}^{*}\delta \hat{a}+i\Omega e^{-i\theta}\delta \hat{b}_{2}
^{\dagger}\nonumber \\
 &  &+\sqrt{2\gamma_{1}} \delta \hat{b}_{1,in}^{\dagger},\nonumber\\
\delta\dot{\hat{\hat{b}}}_{2}^{\dagger} & = & -(\gamma_{2}-i\tilde{\omega}_{2})\delta \hat{b}_{2}^{\dagger}+2iG_{22}^{*}\delta \hat{b}_{2}+iG_{2}^{*}\delta \hat{a}+iG_{2}\delta \hat{a}^{\dagger}
\nonumber \\
 &  &+i\Omega e^{i\theta}\delta \hat{b}_{1}^{\dagger}+\sqrt{2\gamma_{2}}\delta \hat{b}_{2,in}^{\dagger},\nonumber\\
\label{eq:dot1}
\end{eqnarray}
where $\Delta=\Delta_{c}+2g_{1}{\rm }Re[\beta_{1}]+g_{2}(\beta_{2}^{*2}+\beta_{2}^{2}+2\left|\beta_{2}\right|^{2})$ and $\tilde{\omega}_{2}=\omega_{2}+2g_{2}\left|\alpha\right|^{2}$ are the normalized cavity detuning  and the modified mechanical frequency of $b_{2}$, respectively. ${\rm }Re[\beta_{l}]$ extracts the real part of $[\beta_{l}]$. 

Here,  $G_{1}=g_{1}\alpha$ and $G_{2}=4g_{2}\alpha{\rm }Re[\beta_{2}]$ with $G_{22}=g_{2}|\alpha|^2$ represent the linearized optomechanical coupling strengths between the cavity field and the mechanical resonators, respectively. 

To evaluate the cooling performance, we calculate the final average phonon number in both mechanical oscillators. This requires transforming the Langevin equations into a compact form:

\begin{equation}
\mathbf{\dot{u}}(t)=\mathbf{A}\mathbf{u}(t)+\mathbf{N}(t),
\label{eq:du}
\end{equation}
where $\mathbf{u}(t)$ denotes the fluctuation operator vector,
\begin{eqnarray}
\mathbf{u}(t) & = & [\delta \hat{a},\delta \hat{b}_{1},\delta \hat{b}_{2},\delta \hat{a}^{\dagger},\delta \hat{b}_{1}^{\dagger},\delta \hat{b}_{2}^{\dagger}]^{T},
\end{eqnarray}
here, A is the coefficient matrix governing system dynamics, which can be expressed as
\begin{widetext}
\begin{align}
A=\left(\begin{array}{cccccc}
-(\kappa+i\Delta) & -iG_{1} & -iG_{2} & 0 & -iG_{1} & -iG_{2}\\
-iG_{1}^{*} & -(\gamma_{1}+i\omega_{1}) &-i\Omega e^{i\theta}& -iG_{1} & 0 & 0\\
-iG_{2}^{*} &  -i\Omega e^{-i\theta} &-(\gamma_{2}+i\tilde{\omega}_{2}) & -iG_{2} & 0 & -2iG_{22}\\
0 & iG_{1}^{*} & iG_{2}^{*} & -(\kappa-i\Delta) & iG_{1}^{*} & iG_{2}^{*}\\
iG_{1}^{*} & 0 & 0 & iG_{1} & -(\gamma_{1}-i\omega_{1}) &i\Omega e^{-i\theta} \\
iG_{2}^{*} & 0 & 2iG_{22}^{*} & iG_{2} & i\Omega e^{i\theta}& -(\gamma_{2}-i\tilde{\omega}_{2})
\end{array}\right).
\end{align}
\end{widetext}

$\mathbf{N}(t)$ represents the noise operator vector,
\begin{align}
\mathbf{N}(t)&=[\sqrt{2\kappa}\hat{a}_{in},\sqrt{2\gamma_{1}}\hat{b}_{1,in},\sqrt{2\gamma_{2}}\hat{b}_{2,in},\sqrt{2\kappa}\hat{a}_{in}^{\dagger},  \nonumber\\
 &  \sqrt{2\gamma_{1}}\hat{b}_{1,in}^{\dagger}, \sqrt{2\gamma_{2}}\hat{b}_{2,in}^{\dagger}]^{T}.
\end{align}
then the formal solution of the linearized Langevin equation~(\ref{eq:du}) can be written as
\begin{equation}
\mathbf{u}(t)=\mathbf{M}(t)\mathbf{u}(0)+\int_{0}^{t}\mathbf{M}(t-s)\mathbf{N}(s)ds.
\end{equation}
here matrix $\mathbf{M}(t)$ is defined as $\mathbf{M}(t)=\exp(\mathbf{A}t)$.
The steady-state average phonon numbers in two mechanical resonators can be calculated by solving the Lyapunov equation.
It is noteworthy that the parameters used in the subsequent calculations satisfy the stability condition derived from the Routh-Hurwitz criterion, i.e., the real parts of all eigenvalues of the coefficient matrix A are negative.

To study the quantum cooling of the two mechanical resonators, we calculate the final average phonon number by solving the steady-state covariance matrix $\mathbf{V}$. Here, the covariance matrix is constructed for the fluctuation operators defined around each semiclassical steady state, ensuring that the first moments of the original variables are fully incorporated into the linearization parameters. In practice, the Lyapunov equation is solved independently for each stable steady-state branch identified from the nonlinear equations, so that both low- and high-amplitude solutions are properly represented in the results. Covariance matrix  $\mathbf{V}$ is determined by the matrix elements
\begin{eqnarray}
\mathbf{V}_{kl} & = & \frac{1}{2}\left[\left\langle \mathbf{u}_{k}(\infty)\mathbf{u}_{l}(\infty)\right\rangle +\left\langle \mathbf{u}_{l}(\infty)\mathbf{u}_{k}(\infty)\right\rangle \right](k,l=1\sim6).\nonumber\\
\end{eqnarray}

Under the stability condition, the steady-state covariance matrix
\textbf{$\mathbf{V}$} meets the Lyapunov equation,
\begin{equation}
\mathbf{A}\mathbf{V}+\mathbf{V}\mathbf{A}^{T}+\mathbf{Q}=0,
\label{eq:liyapu}
\end{equation}
where $T$ represents the matrix transpose operation and the matrix \textbf{$\mathbf{Q}$} is given by
\begin{equation}
\mathbf{Q}=\frac{1}{2}(\mathbf{C}+\mathbf{C}^{T}),
\end{equation}
with $\mathbf{C}$ representing the noise correlation matrix characterized by the
matrix elements
\begin{equation}
\left\langle \mathbf{N}_{k}(s)\mathbf{N}_{l}(s')\right\rangle =\mathbf{C}_{k,l}\delta(s-s'),
\end{equation}
where $s  =s' \longrightarrow \left\langle \mathbf{N}_{k}(s)\mathbf{N}_{l}(s')\right\rangle =\mathbf{C}_{k,l}$ and
$s  \neq s'\longrightarrow \left\langle \mathbf{N}_{k}(s)\mathbf{N}_{l}(s')\right\rangle =0$.
The constant matrix $\mathbf{C}$, with the Markovian bath is considered in this study, is expressed as follows:
\begin{eqnarray}
\mathbf{C}=\left(\begin{array}{cccccc}
0 & 0 & 0 & 2\kappa & 0 & 0\\
0 & 0 & 0 & 0 & 2\gamma_{1}(\bar{n}_{1}+1) & 0\\
0 & 0 & 0 & 0 & 0 & 2\gamma_{2}(\bar{n}_{2}+1)\\
0 & 0 & 0 & 0 & 0 & 0\\
0 & 2\gamma_{1}\bar{n}_{1} & 0 & 0 & 0 & 0\\
0 & 0 & 2\gamma_{2}\bar{n}_{2} & 0 & 0 & 0
\end{array}\right).\nonumber\\
\end{eqnarray}

Under these conditions, we solve the Lyapunov Eq.~(\ref{eq:liyapu}) to determine the covariance matrix $\mathbf{V}$, which yields the final steady-state phonon numbers:
\begin{eqnarray}
n_{1,f}=\left\langle \delta \hat{b}_{1}^{\dagger}\delta \hat{b}_{1}\right\rangle =\mathbf{V}_{52}-\frac{1}{2},\nonumber\\
n_{2,f}=\left\langle \delta \hat{b}_{2}^{\dagger}\delta \hat{b}_{2}\right\rangle =\mathbf{V}_{63}-\frac{1}{2},
\end{eqnarray}
where $\mathbf{V}_{52}$ and $\mathbf{V}_{63}$ are derived through the solution of the Lyapunov equation.

\begin{figure}[ht]
\includegraphics[width=0.5\textwidth]{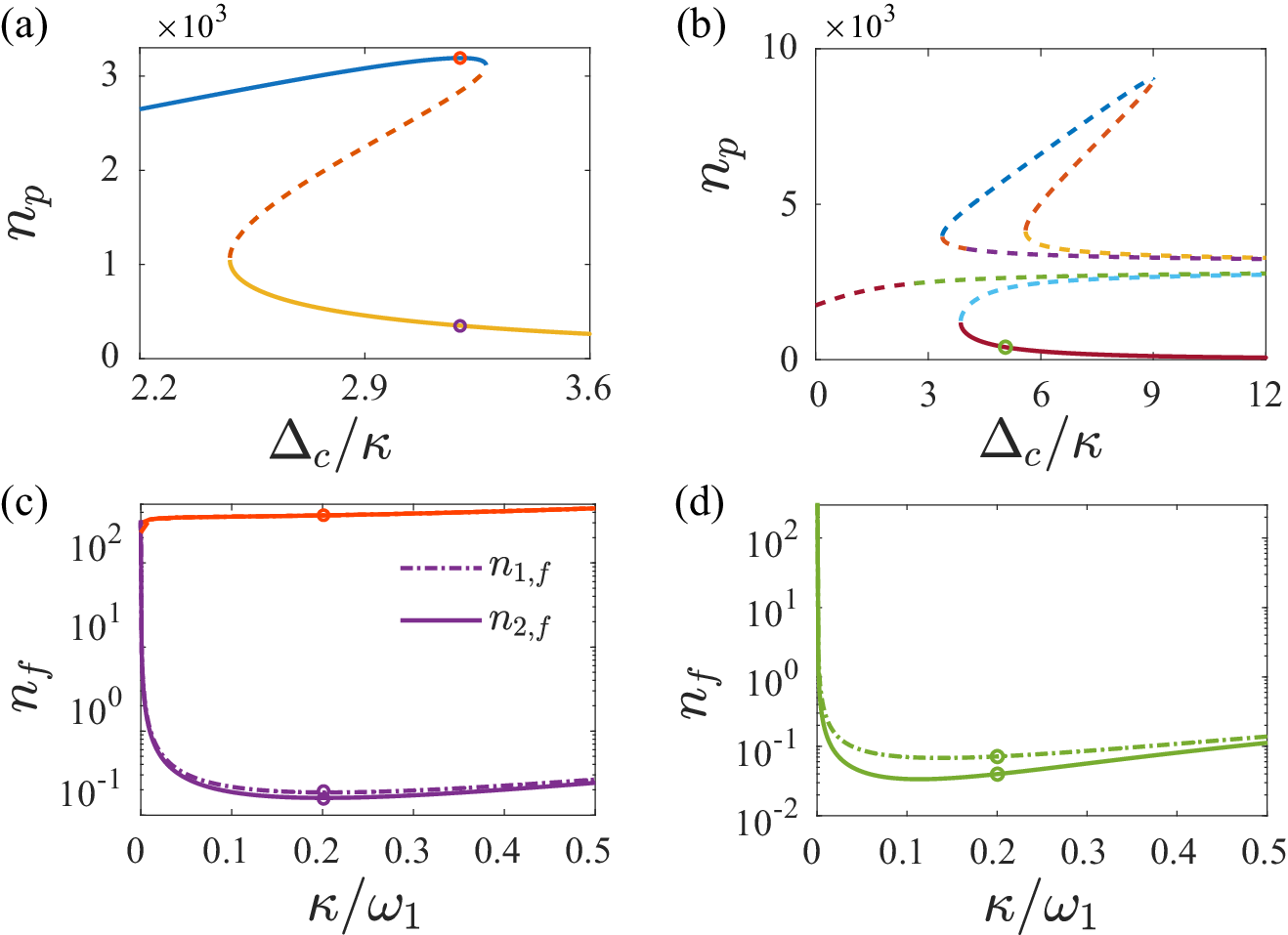}
\caption{(a) Steady-state intracavity photon number $n_p$ versus cavity detuning $\Delta_c$ without quadratic coupling ($g_2=0$), showing conventional bistability ($g_1/\kappa=0.05$, $\omega_1/\kappa=\omega_2/\kappa=5$, $\eta/\kappa=56.5$, $\Omega/\kappa=0.2$).  The marked detuning point ($\Delta_c/\kappa=3.2$) is used for cooling analysis. (b) $n_p$ versus $\Delta_c/\kappa$ in the presence of quadratic coupling ($g_1/\kappa=0.05$, $g_2/\kappa=-0.0004$, $\eta/\kappa=95$, $\Omega/\kappa=1$, and $\omega_1/\kappa=\omega_2/\kappa=5$).  The marked detuning point ($\Delta_c/\kappa=5$) is used for cooling analysis. (c,d) Final phonon occupations $n_{1,f}$ and $n_{2,f}$ as functions of $\kappa/\omega_1$, evaluated at the dynamically stable steady states corresponding to the marked detuning points in (a) and (b), respectively.}
\label{fig:c}
\end{figure}

To establish the connection between the nonlinear steady-state structure and mechanical ground-state cooling, we first consider the purely linear optomechanical case ($g_2=0$), as shown in Fig.~\ref{fig:c}(a). In this regime, the system exhibits genuine bistability, characterized by two dynamically stable steady-state branches separated by an unstable one. Only the stable branches correspond to physically accessible operating points. When quadratic optomechanical coupling is introduced [Fig.~\ref{fig:c}(b)], the nonlinear steady-state equation admits up to five or seven real algebraic solutions depending on the detuning. However, linear stability analysis based on the eigenvalues of the Jacobian matrix reveals that, in the parameter regime considered, typically only the lowest branch remains dynamically stable, while the higher branches are unstable and thus not physically realizable steady states. Consequently, cooling analysis in the quadratic-coupling regime is performed exclusively around this dynamically stable branch. Unstable solutions are not used in the linearization or cooling evaluation.
We next examine the corresponding cooling behavior, as presented in Figs.~\ref{fig:c}(c) and~\ref{fig:c}(d). These panels show the final phonon occupations $n_{1,f}$ and $n_{2,f}$ as functions of $\kappa/\omega_1$, evaluated at the steady-state points marked by purple, red, and green circles in Figs.~\ref{fig:c}(a) and \ref{fig:c}(b), respectively. 
We emphasize that all cooling results shown in Figs.~\ref{fig:c}(c) and \ref{fig:c}(d) are obtained from dynamically stable steady-state solutions. For each value of $\kappa/\omega_1$, stability is verified by ensuring that all eigenvalues of the corresponding Jacobian matrix have negative real parts. As $\kappa/\omega_1$ increases, both $n_{1,f}$ and $n_{2,f}$ initially decrease, reach a minimum, and then gradually saturate. This non-monotonic behavior indicates the existence of an optimal cavity decay rate that minimizes the mechanical energy. Physically, this reflects a balance between intracavity photon number and cavity dissipation: a moderate decay rate allows strong optomechanical interaction for efficient cooling while simultaneously enabling effective dissipation of excess thermal noise.

Importantly, we find that both mechanical resonators can be cooled close to their quantum ground states on the dynamically stable branch. For example, in the purely linear bistable regime [Fig. \ref{fig:c}(c)], the lowest stable branch yields $n_{1,f}=0.09$ and $n_{2,f}=0.08$ when $n_p=347$, as marked by the purple circles, indicating near-ground-state cooling. Moreover, we find that the mechanical resonators on the higher stable branches (e.g., $n_p=3191$, red lines in Fig. \ref{fig:c}(c)) do not exhibit efficient ground-state cooling. This can be understood from two main aspects. First, the large intracavity photon number associated with the upper branch enhances nonlinear optomechanical effects, which reduces the effectiveness of the linearized cooling mechanism in this regime. Second, the strong nonlinearity induces a significant shift of the effective cavity detuning, causing it to deviate from the optimal red-sideband condition, thereby suppressing the cooling process. As a result, the lowest branch represents the preferred operating regime for achieving dual-mode ground-state cooling.
In contrast, in the presence of quadratic coupling [Fig. \ref{fig:c}(d)], when $n_p=410$ (green circles), cooling performance improves further, reaching $n_{1,f}=0.07$ and $n_{2,f}=0.04$. These results demonstrate that the presence of a higher-order nonlinear steady-state structure does not obstruct cooling. Even though multiple algebraic steady-state solutions may exist, efficient dual-mode ground-state cooling is achieved on the dynamically stable branch. The improvement in the quadratic-coupling regime originates from the modification of the nonlinear steady-state structure, which reshapes the effective linearized parameters governing the cooling dynamics. 

These results reveal a more general connection between nonlinear multistability and dual-mode cooling. The nonlinear steady-state equations admit multiple branches characterized by different intracavity photon numbers. Each dynamically stable branch therefore defines a distinct operating point of the system, with different effective detuning and effective optomechanical couplings. Since these effective parameters directly govern the optical damping and the final phonon occupations, different stable branches naturally lead to different cooling efficiencies. In this sense, nonlinear multistability is not merely a mathematical property of the steady-state equation, but provides multiple physically accessible cooling regimes within the same system, where the selected dynamically stable branch determines the achievable cooling performance.

To clarify the physical mechanism, we note that each steady-state branch corresponds to a distinct intracavity field amplitude $ \alpha $ (or photon number $ n_p = |\alpha|^2 $). Consequently, the effective optomechanical couplings and detuning differ among branches and directly determine the cooling rate. In particular, the coupling strengths $G_1 = g_1 \alpha$ and $G_2 = 4g_2 \alpha \text{Re}[\beta_2]$, along with the photon-number-dependent detuning $\Delta$, govern the optical damping rate. The collective cooperativity, which scales with the square of the effective couplings, is a key factor. Branches with larger $n_p$ generally exhibit stronger $G_1$ and $G_2$, and the effective detuning $\Delta$ may move the system closer to the optimal sideband condition $\Delta \approx -\omega_m$, thereby reducing the final phonon occupation. In contrast, branches with weaker intracavity fields or detunings far from resonance yield smaller effective cooperativity and inferior cooling performance.

Moreover, the interplay between linear and quadratic coupling pathways can induce interference effects analogous to dark modes, which partially decouple the mechanical motion from the cavity field and suppress cooling. Superior cooling therefore arises when the steady-state parameters simultaneously provide strong cooperativity, near-optimal detuning, and suppression of destructive interference. Here, nonlinear steady-state restructuring refers to the modification of the semiclassical steady-state solutions induced by nonlinear optomechanical interactions, which leads to different intracavity photon numbers and consequently alters the effective detuning and optomechanical coupling strengths governing the cooling dynamics. Overall, these results establish a clear physical connection between the nonlinear steady-state structure and mechanical ground-state cooling. Rather than being a purely mathematical feature, this steady-state structure provides a mechanism for tuning the effective cooling parameters and optimizing dual-mode cooling performance in multimode optomechanical systems.

\section{Enhanced Ground state cooling of two mechanical resonators}
\begin{figure}[t]
\includegraphics[width=0.49\textwidth]{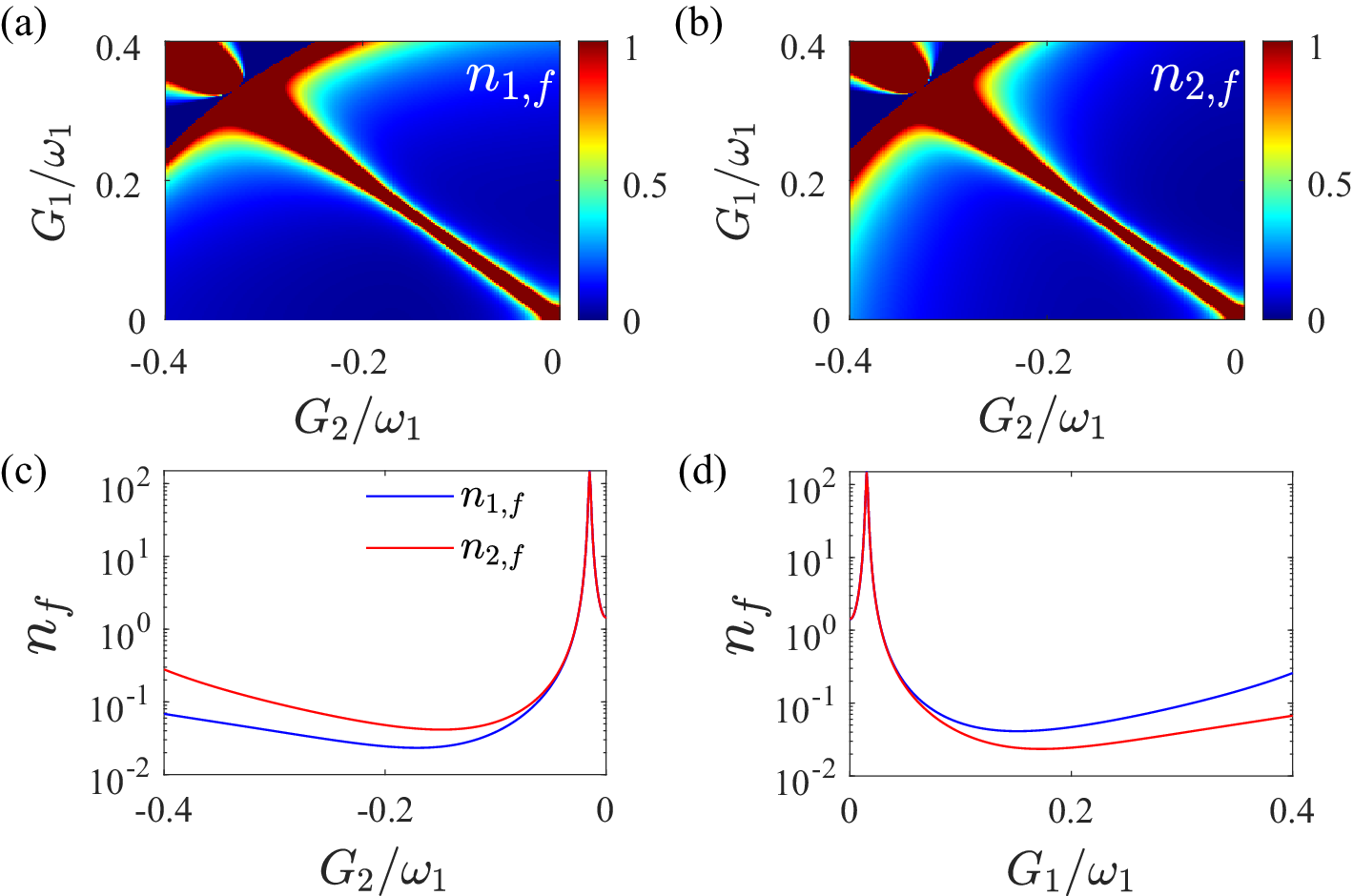}
\caption{Final average phonon numbers (a) $n_{1,f}$ and (b) $n_{2,f}$ on parameter plane $G_1-G_2$. (c) $ n_f $ as a function of $G_2/\omega_1$ with $G_1/\omega_1 = 0.015$. (d) $ n_f $ as a function of $G_1/\omega_1$ with $G_2/\omega_1 = -0.015$.
The other parameters are set as $\Delta/\omega_1=\tilde{\omega} _{2}/\omega_1=1$, $G_{22}/\omega_1=-0.01$, $\theta=\pi$, $\kappa=0.1$, $\Omega/\omega_1=0.13$, $\gamma_{1}/\omega_1=\gamma_{2}/\omega_1=2\times10^{-6}$ and $\bar{n}_{1}=\bar{n}_{2}=300$.}
\label{fig:6}
\end{figure}
Motivated by the observation that efficient dual-mode cooling can be achieved on the dynamically stable branch of the nonlinear steady-state structure, we now shift our focus from specific operating points to a more general analysis of the cooling mechanism. Rather than emphasizing the multiplicity of steady-state solutions, our goal here is to identify the fundamental parameter conditions that robustly enable simultaneous ground-state cooling over a broad range of system parameters. To this end, we analyze the linearized dynamics around a stable semiclassical steady state and extract the essential coupling structure responsible for cooling. Under the rotating-wave approximation (RWA) and at the optimal red-detuning condition $\Delta \approx \omega_{1}\approx \tilde{\omega}_{2}$,  the linearized Hamiltonian (neglecting noise terms) takes the form:
\begin{eqnarray}
\hat{H}_{\rm RWA} & =  \Delta\delta \hat{a}^{\dagger}\delta \hat{a}+\omega_{1}\delta \hat{b}_{1}^{\dagger}\delta \hat{b}_{1}+\tilde{\omega}_{2}\delta \hat{b}_{2}^{\dagger}\delta \hat{b}_{2}
+[(G_{1}^{*}\delta \hat{a}\delta \hat{b}_{1}^{\dagger} \nonumber\\
&+  \!G_{2}^{*}\delta \hat{a}\delta \hat{b}_{2}^{\dagger} \!+ \! \Omega e^{i\theta}\delta \hat{b}_{1}^{\dagger}\delta \hat{b}_{2} \!+\! {2G_{22}\delta\hat{b}_{2}^{\dagger 2}}) \!+\! \rm H.c.].
\end{eqnarray} 

In this expression, rapidly oscillating non-resonant terms are omitted as they contribute negligibly to the cooling dynamics. The effective couplings $G_1$ and $G_2$ enable energy exchange between the cavity and the two mechanical modes, while the direct phonon-exchange interaction $\Omega$ and the quadratic-induced term $G_{22}$ introduce additional coupling pathways unique to the present system.

To further elucidate the physical origin of the interference mechanism discussed above, we analyze the structure of the effective linearized interactions. The optical fluctuation mode mediates two coupling channels characterized by $G_1$ and $G_2$, while the mechanical modes are also coupled via $\Omega e^{i\theta}$ and $G_{22}$. For clarity, neglecting $\Omega$ and $G_{22}$, the interaction reduces to $G_1 \,\delta \hat a\, \delta \hat b_1^\dagger + G_2 \,\delta \hat a\, \delta \hat b_2^\dagger + \mathrm{H.c.}$, which couples the cavity only to the bright mode $\delta \hat b_B \propto G_1 \delta \hat b_1 + G_2 \delta \hat b_2$,  while the orthogonal dark mode $\delta \hat b_D \propto G_2 \delta \hat b_1 - G_1 \delta \hat b_2$ is decoupled from the cavity. This indicates that cooling is suppressed near $G_1 \approx G_2$, where destructive interference inhibits efficient energy extraction. When finite $\Omega$ and/or $G_{22}$ are included, the bright and dark modes become mixed, lifting the destructive interference and restoring simultaneous cooling of both mechanical modes. 

With this physical understanding of the interference mechanism, we now explore the parameter space and the resulting cooling behavior. Here, the effective couplings $G_{1,2}$ are treated as parameters within the linearized model to elucidate the underlying cooling mechanism. In the full nonlinear system, however, $G_{1,2} \propto \alpha$ are determined self-consistently by the steady-state cavity field and correspond to different dynamically stable branches.

Figs. \ref{fig:6} (a) and \ref{fig:6}(b) show the final average phonon numbers $n_{1,f}$ and $n_{2,f}$  as functions of the linear ($G_1$) and quadratic ($G_2$) optomechanical coupling strengths. Clearly, simultaneous ground-state cooling of both mechanical oscillators is achievable over a wide range of parameters, as indicated by the blue-shaded areas.  However, when the linear and quadratic couplings become comparable ($G_1 \approx |G_2|$), the dark-mode effect emerges and suppresses effective cooling, preventing both oscillators from reaching their ground states. Notably, the phonon-exchange interaction introduces a cooling asymmetry: dominance of $G_1$ preferentially enhances the cooling of $\hat{b}_2$, whereas dominance of $G_2$ leads to a lower phonon occupation in $\hat{b}_1$.

To gain deeper insight into this asymmetric behavior, we analyze how the final phonon numbers respond to the variation of each coupling parameter individually.  Fig. \ref{fig:6}(c) shows the dependence of the final phonon numbers on the quadratic coupling strength $G_2$, with the linear coupling fixed at $G_1/\omega_1=0.015$.  In the absence of quadratic coupling, i.e., when only linear interaction is present, the mechanical resonators cannot reach their ground states. As $G_2$ increases, the phonon numbers first rise, then drop to a minimum, and finally rise gently. When $| G_2|/\omega_1>0.028$, effective cooling is realized, and oscillator ${b}_1$ shows significantly better cooling performance than ${b}_2$. A similar trend is observed in Fig.~\ref{fig:6}(d), which displays the variation of phonon numbers with linear coupling $G_1$ while keeping $G_2/\omega_1=-0.015$ constant.  In the case of only quadratic coupling, ground-state cooling is again unattainable.  When $G_1/\omega_1> 0.028$, ${b}_2$ achieves better cooling than ${b}_1$, indicating the reverse asymmetry compared to the  $G_2$-dominant case.

Furthermore, our results reveal that simultaneous ground-state cooling of both mechanical oscillators is achievable over a wide range of parameters, except in the symmetric region where $G_1 \approx |G_2|$. In this special case, the dark-mode effect severely suppresses energy dissipation, leading to cooling failure. This phenomenon is consistent with earlier findings on dark-mode suppression via phase-control techniques~\cite{PhysRevA.102.011502}.  To overcome this obstacle and restore efficient joint cooling, we further investigate strategies to break the dark-mode coherence. In particular, we explore the role of the second-order optomechanical-induced frequency shift $G_{22}$, which introduces an effective energy detuning capable of disrupting the destructive interference associated with dark modes.

\begin{figure}[t]
\includegraphics[width=0.49\textwidth]{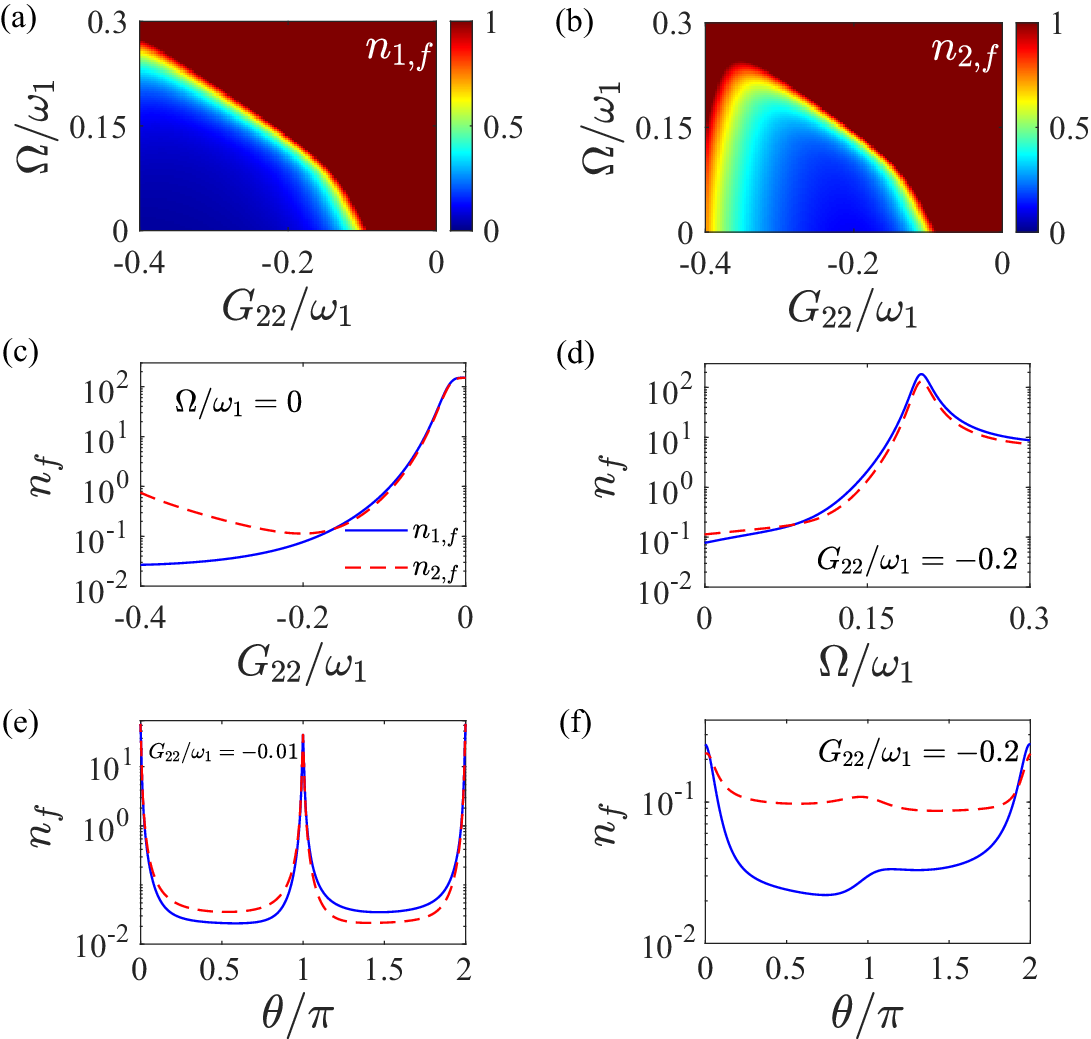}
\caption{Final average phonon numbers $n_{1,f}$  (a)  and $n_{2,f}$  (b) on the $G_{22}$-$\Omega$ plane with $\theta=\pi$.  (c) Final average phonon number \( n_f \) as a function of  $( G_{22}/\omega_1 )$ with $\Omega/\omega_1 = 0$ and $\theta=\pi$. (d) Final average phonon number \( n_f \) as a function of  \( \Omega/\omega_1 \) with $G_{22}/\omega_1=-0.2$ and $\theta=\pi$. (e, f) Dependence of the final average phonon number \( n_f \) on the normalized phase \( \theta \) for \( \Omega/\omega_1 = 0.1 \), with $G_{22}/\omega_1 = -0.01 $ and $ -0.2$. The other parameters are $G_1/\omega_1=0.1$, $G_2/\omega_1=-0.1$, $\Delta/\omega_1=\tilde{\omega}_{2}/\omega_1=1$, $\kappa/\omega_1=0.1$, $\gamma_{1}/\omega_1=\gamma_{2}/\omega_1=2\times10^{-6}$, and $\bar{n}_{1}=\bar{n}_{2}=300$.}
\label{fig:6g22}
\end{figure}

Figs.~\ref{fig:6g22}(a) and ~\ref{fig:6g22}(b) present the final average phonon numbers $n_{1,f}$ and $n_{2,f}$ as functions of the second-order optomechanical-induced frequency shift $G_{22}$ and the phonon exchange strength $\Omega$. It is evident that when $|G_{22}|$ is small, simultaneous ground-state cooling of both mechanical oscillators cannot be achieved, as the system remains trapped in a dark-mode-dominated regime that inhibits energy dissipation. However,  as $|G_{22}|$ increases, an effective energy detuning is introduced, breaking the destructive interference associated with the dark mode. This disruption of coherence opens new dissipative channels, thereby enabling simultaneous ground-state cooling in regions where the phonon exchange interaction remains weak. 
These results indicate that different dynamically stable steady-state branches correspond to distinct effective operating points, which in turn lead to different cooling efficiencies.

To elucidate this mechanism further, Fig.~\ref{fig:6g22}(c) examines the variation of the phonon numbers as a function of $G_{22}$ in the absence of phonon exchange ($\Omega=0$).  As $|G_{22}|$ increases, the phonon number of oscillator $\hat{b}_1$ decreases monotonically, reaching the ground-state cooling regime.  In contrast,  $\hat{b}_2$ exhibits a non-monotonic behavior, featuring a pronounced dip with $n_{2,f}=0.11$, which corresponds to a parameter window where both oscillators can be simultaneously cooled to their ground states. This behavior highlights the distinct role of $G_{22}$ in selectively modulating the cooling pathways of the two oscillators through an interplay between frequency detuning and dissipative coupling.

Fig.~\ref{fig:6g22}(d) explores the influence of phonon exchange by fixing $G_{22}/\omega_1 = -0.2$ and varying $\Omega$. For small values of $\Omega$, both mechanical oscillators achieve ground-state cooling. However, as $\Omega$ increases, the phonon numbers of both oscillators rise gradually. This degradation of cooling performance is attributed to the re-emergence of phonon-mediated coherent coupling, which effectively counteracts the dissipative advantage introduced by $G_{22}$ and partially restores dark-mode-like interference effects. These results underscore that appropriately tuning $G_{22}$ provides a powerful means to break the dark-mode coherence and optimize the cooling of multi-mode optomechanical systems. Conversely, excessive phonon exchange can reintroduce coherence and suppress cooling, thereby revealing a delicate balance between dissipative and coherent interactions. This insight offers valuable guidance for the design and cooperative control of complex multi-mode quantum systems where precise management of mode hybridization and dissipation is critical.

The case of $\theta=\pi$ is selected as it corresponds to the condition where the destructive interference between the two optomechanical channels is strongest, making the dark-mode effect most evident. From the collective-mode perspective, the dark mode emerges when the effective couplings satisfy $G_1+G_2e^{i\theta}$=0, which decouples one collective mechanical mode from the cavity field and suppresses cooling. Furthermore, the phase-scan results in Figs.~\ref{fig:6g22}(e) and \ref{fig:6g22}(f) illustrate that tuning the phase $\theta$ effectively breaks this dark mode, decoupling photons from mechanical excitations and enabling efficient ground-state cooling. Once the dark mode is lifted, the phonon number can be reduced below unity. Importantly, the frequency shift and phase control act synergistically, allowing both mechanical oscillators to reach the ground-state regime rather than only one. These results confirm that the proposed cooling mechanism remains robust even for $\theta \neq \pi$, demonstrating its broad applicability.

\begin{figure}[t]
\includegraphics[width=0.49\textwidth]{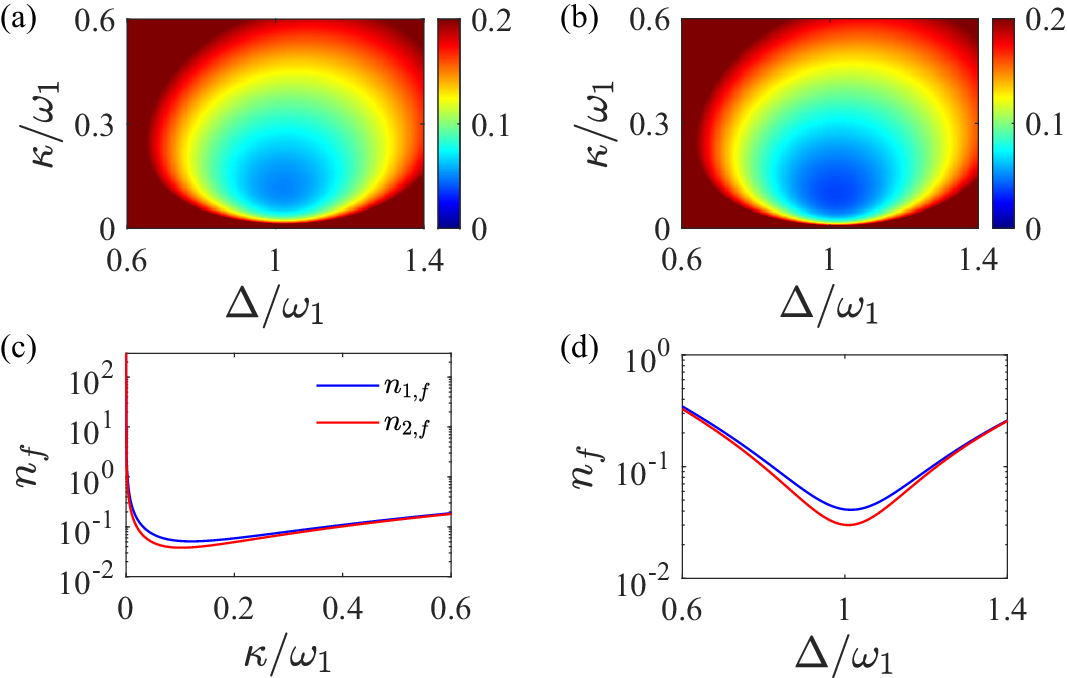}
\caption{Final average number of phonon (a) $n_{1,f}$ and (b) $n_{2,f}$ on the parameter plane $\Delta-\kappa$. (c) Final average phonon number \( n_f \)  at the red detuning sideband with $\Delta/\omega_1=\tilde{\omega} _{2}/\omega_1=1$. (d) Final average phonon number \( n_f \) at $\kappa/\omega_1=0.1$. The other parameters are $G_1/\omega_1=0.1$, $G_2/\omega_1=-0.01$, $G_{22}/\omega_1=-0.01$, $\Omega/\omega_1 = 0.1$, $\Delta/\omega_1=\tilde{\omega}_{2}/\omega_1=1$, $\theta=\pi$, $\kappa/\omega_1=0.1$, $\gamma_{1}/\omega_1=\gamma_{2}/\omega_1=2\times10^{-6}$
and $\bar{n}_{1}=\bar{n}_{2}=300$.}
\label{fig:c1}
\end{figure}

To foster a deeper understanding of the ground-state cooling of mechanical oscillators, {Figs.~\ref{fig:c1}(a) and \ref{fig:c1}(b) depict the phonon numbers $n_{1,f}$ and $n_{2,f}$ as functions of the cavity detuning $\Delta/\omega_1$ and the cavity decay rate $\kappa/\omega_1$. Evidently, under red-detuning conditions where $\Delta = \omega_1 = \tilde{\omega}_2$ and and in the regime of low cavity dissipation, both mechanical modes can reach low-phonon steady states, enabling efficient ground-state cooling. To further elucidate the role of cavity dissipation, Fig.~\ref{fig:c1}(c) shows the the dependence of $n_{1,f}$ and $n_{2,f}$ on $\kappa$.  As $\kappa$ increases, the phonon numbers initially drop rapidly below unity and subsequently saturate, confirming that ground-state cooling is successfully achieved for both mechanical oscillators. The optimal cooling performance
is characterized by $n_{1,f} = 0.045$ and $n_{2,f} = 0.035$ at $\Delta/\omega_1 = 1$, as illustrated in Fig. \ref{fig:c1}(d), demonstrating that cavity dissipation plays a pivotal role in controlling the mechanical energy.

This behavior can be understood in terms of the competition between optomechanical coupling strength and cavity dissipation. At low $\kappa$, the system operates in the resolved-sideband regime, where sideband cooling becomes efficient by suppressing the Stokes (heating) process while enhancing the anti-Stokes (cooling) transition. Moreover, the dependence of phonon number on cavity detuning $\Delta$ also highlights the crucial influence of photon-mediated backaction. Optimal cooling occurs near the red sideband ($\Delta \approx \omega_m$), where the anti-Stokes process dominates. Deviations from this detuning condition reduce the cooling efficiency due to increased Stokes scattering, which adds energy back to the mechanical modes. These results collectively underscore the importance of  both the cavity dissipation rate and the driving detuning in realizing efficient ground-state cooling. They also provide a practical framework for engineering low-phonon steady states in multimode optomechanical systems, particularly when multiple types of mechanical modes (e.g., linearly and quadratically coupled) are involved.

\section{Conclusion}

In summary, we have theoretically explored an optomechanical system incorporating both linear and quadratic coupling between a single-mode optical cavity and two mechanical resonators.  By systematically tuning the optical cavity detuning, the driving strength, and phonon-exchange interactions, we demonstrated the emergence of rich optical bistability and multistability behavior, with up to seven steady-state solutions induced by the quadratic optomechanical coupling. A detailed stability analysis reveals that only specific branches are dynamically stable. Crucially, we showed that ground-state cooling of both mechanical oscillators can be simultaneously realized  on the dynamically stable branch within the nonlinear regime, thereby enabling quantum control of mechanical motion under nonlinear dynamical regimes. Furthermore, we conducted a systematic study of dual-mode ground-state cooling beyond the multistability regime. In this general analysis, we demonstrated that robust and controllable simultaneous ground-state cooling can be realized by engineering the optomechanical coupling asymmetry, tuning second-order frequency shifts to break dark-mode interference. This mechanism opens new dissipative channels and enhances cooling efficiency. We also examined the influence of cavity detuning and dissipation, providing valuable insights into the cooperative control of multi-mode optomechanical systems. Our findings establish a versatile framework for achieving tunable nonlinear optical responses and robust quantum control of multimode optomechanical systems. The results pave the way for future developments of multifunctional quantum devices based on quadratic optomechanical interactions, with potential applications in nonclassical quantum switching, nonlinear optics, scalable quantum memories, and multimode quantum information processing.

{\em Acknowledgments}.---This work was supported by the National Natural Science Foundation of China (Grant No.12374365, Grant No. 12274473, and Grant No. 12135018), Quantum Science and Technology-National Science and Technology Major Project (Grant No.2025ZD0300400), Guangdong Provincial Quantum Science Strategic Initiative (Grant No. GDZX2505001), and Guangdong University of Technology SPOE Seed Foundation (SF2024111504).  

%%
%%\bibliographystyle{unsrt}
%\bibliography{quadradic}

%

\end{document}